# A Practical Approach of Actions for FAIRification Workflows

Natalia Queiroz de Oliveira[1][0000-0001-8371-142X], Vânia Borges[1][0000-0002-6717-1168], Henrique F. Rodrigues[1][0000-0002-6777-3935], Maria Luiza Machado Campos[1][0000-0002-7930-612X], Giseli Rabello Lopes[1][0000-0003-2482-1826]

[1] Federal University of Rio de Janeiro, Rio de Janeiro, Brazil

natalia.oliveira@ppgi.ufrj.br, vjborges30@ufrj.br,
hfr@ufrj.br, mluiza@ppgi.ufrj.br, giseli@ic.ufrj.br

**Abstract** Since their proposal in 2016, the FAIR principles have been largely discussed by different communities and initiatives involved in the development of infrastructures to enhance support for data findability, accessibility, interoperability, and reuse. One of the challenges in implementing these principles lies in defining a well-delimited process with organized and detailed actions. This paper presents a workflow of actions that is being adopted in the VODAN BR pilot for generating FAIR (meta)data for COVID-19 research. It provides the understanding of each step of the process, establishing their contribution. In this work, we also evaluate potential tools to (semi)automatize (meta)data treatment whenever possible. Although defined for a particular use case, it is expected that this workflow can be applied for other epidemical research and in other domains, benefiting the entire scientific community.

## 1 Introduction

Since its publication in 2016 [1], the FAIR principles have been guiding best practices on publishing scientific research data and their associated metadata to make them Findable, Accessible, Interoperable, and Reusable by humans and especially by machines. The international GO FAIR[1] initiative aims at implementing the FAIR principles through Implementation Networks (INs)[2], which operate as FAIR drivers, collaboratively involving communities, institutions, and countries. As a sense of urgency due to the rapidly COVID-19 pandemic spread, the Data Together initiative involving the Committee on Data for Science and Technology (CODATA)[3], Research Data Alliance (RDA)[4], World Data System (WDS)[5], and GO FAIR was established, including a joint effort, the Virus Outbreak Data Network IN (VODAN-IN[6]). The initial goal is to develop a federated data infrastructure to support the capture and use of data related to epidemic outbreaks, both for the current situation and future epidemics.

---

[1] https://www.go-fair.org/
[2] https://www.go-fair.org/implementation-networks/
[3] https://codata.org/

Initiatives such as VODAN-IN attempt to deliver FAIR data, in the original sense of the acronym, but also in the sense of "Federated, AI-Ready"[7] data, therefore readable and machine actionable. The first IN of GO FAIR Brazil [2], GO FAIR Brazil Health[8], is a thematic network responsible for developing strategies for the implementation of the FAIR principles in the health domain. VODAN BR[9] is the first pilot of GO FAIR Brazil Health, aiming to collect and implement a data management infrastructure for COVID-19 hospitalized patients' cases, according to the FAIR principles. The published scientific research data and their associated metadata should be as open as possible and as closed as necessary, to protect participant privacy and reduce the risk of data misuse.

The attempt of adopting the FAIR principles has led many scientific disciplines, which value the importance of research data stewardship [3], to consider: "Which knowledge is needed to make my data FAIR?" or "What solutions could be used?". The process of making data FAIR is called FAIRification and the VODAN BR pilot has been using the FAIRification workflow [4] for the transformation and publication of FAIR (meta)data. This general workflow describes a process that applies to any type of data and can be extended, adapted, and reused in different domains. However, in the VODAN BR pilot, we verified the need for specific actions to be defined in a more detailed FAIRification process, as a basis for implementation choices that needed to be carried out to support it.

Based on the recommendations of the original FAIRification process, this paper presents a practical approach for actions associated with the transformation of data and metadata, which are being tested in the VODAN BR pilot to ensure the publication of FAIR (meta)data on COVID-19. To systematize some of the established actions, we experimented and analyzed potential solutions to support FAIRification.

The remainder of this paper is organized as follows: Section 2 presents an overview of the FAIRification workflow; Section 3 describes the actions established for each step of the workflow for the VODAN BR pilot; Section 4 presents support solutions analyzed during the study of the steps, aiming at the systematization of the process; Section 5 presents a discussion about the relevant aspects treated in this work and concludes with final comments for future work.

## 2    FAIRification Workflow

---

[4] https://rd-alliance.org/
[5] https://world-datasystem.org/
[6] https://go-fair.org/wp-content/uploads/2020/03/Data-Together-COVID-19-Statement-FINAL.pdf
[7] https://www.go-fair.org/implementation-networks/overview/vodan/
[8] https://www.go-fair-brasil.org/saude
[9] https://portal.fiocruz.br/en/vodan-brazil

The generic workflow proposed in [4] aims to facilitate the FAIRification process comprising three defined phases: pre-FAIRification, FAIRification and post-FAIRification. The phases are further divided into seven steps: 1) identify the FAIRification objective; 2) analyze data; 3) analyze metadata; 4) define semantic model for data (4a) and metadata (4b); 5) make data (5a) and metadata (5b) linkable; 6) host FAIR data; and 7) assess FAIR data.

From these multiple steps, the authors describe how data and metadata can be processed, which knowledge is required, and which procedures and tools can be used to obtain FAIR (meta)data. The FAIRification workflow was defined based on discussions and experimentations from a series of workshops (Bring your own device - BYOD) [5] and is applicable to any kind of data and metadata.

FAIRification is in fact a complex process, requiring several areas of expertise and data stewardship knowledge. Our adaptation follows the steps of the generic FAIRification workflow and, according to our understanding, steps 6 and 7 have been renamed to 6) host FAIR data and metadata and 7) assess FAIR data and metadata. The reason is to emphasize the importance of storing, publishing, and evaluating both FAIR data and metadata.

In the literature review we found related studies and experiments discussing the FAIRification process. In [6], a retrospective form of FAIRification approach is presented, using two related metabolic datasets associated with journal articles to curate and re-annotate data and metadata using interoperability standards. However, the work does not follow the generic FAIRification workflow approach.

The work of [7] details the FAIRification process[10] proposed by GO FAIR, which aims to facilitate the conversion of spreadsheets into FAIR databases, with the help of the NMDataParser tool [8]. This tool supports data aggregation block levels, developed to speed up the mapping of the original file into the eNanoMapper[11] semantic model.

In [9], the authors present an architecture, following the GO FAIR FAIRification process, and addressing identified gaps in the process when dealing with datasets from the health domain. Another paper [10] proposes the De-novo FAIRification method, based on an Electronic Data Capture (EDC) system, where the steps of the generic FAIRification workflow are incorporated into the data collection process for a registration or research project.

We verified that these related works present approaches with guidelines for FAIRification proposed by the generic workflow and for the FAIRification process of GO FAIR. However, none of the works present the detail of associated actions for the FAIRification in a delimited and specific way, justifying implementation choices to support the transformation and publishing of FAIR (meta)data.

## 3 Set of Actions for FAIRification

### 3.1 VODAN BR Pilot

---

[10] https://www.go-fair.org/fair-principles/fairification-process/
[11] https://search.data.enanomapper.net/

The VODAN BR pilot has been using the adapted FAIRification workflow, in a platform acting as a FAIR solution for COVID-19 clinical data. This platform is not only concerned with the process of data transformation and metadata generation, but also with support solutions to host and publish FAIR (meta)data. Figure 1 shows a diagram of the platform, with the (meta)data flow from the original source to the target FAIR Data Point, associated with the steps of the adapted FAIRification workflow.

For the pilot, the platform captures COVID-19 patients' datasets, in CSV format (1), applying the pre-FAIRification phase (a) steps. This dataset and the results of the performed analyses are used in steps 4a and 4b of the FAIRification phase (b), establishing the semantic models. Following the actions specified by steps 5a and 5b, the Extract, Transform and Load (ETL) process, designated in this work as ETL4FAIR (2), is responsible for transforming data and metadata to the RDF representation.

Hosting of (meta)data follows step 6, with the linkable (meta)data published in a triplestore (3). A triplestore Application Programming Interface (API) can be made available for access to (meta)data. The metadata schemas for the dataset and its distributions are provided in a FAIR Data Point (5). The distribution metadata schemas can provide an URL to download the RDF file published in the repository (4) and/or an SPARQL endpoint to the triplestore.

Finally, step 7, intended for the assessment of FAIR (meta)data, in the post-FAIRification phase (c), allows the (meta)data FAIRness evaluation and verifying the suitability of the established platform.

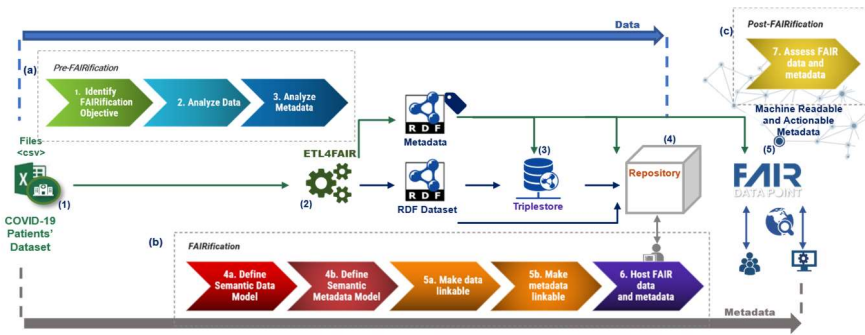

**Fig. 1.** VODAN BR pilot platform, with associated steps of the adapted FAIRification workflow.

### 3.2 Phases, Steps and Actions

In the platform, we transformed the recommendations from each step of the adapted FAIRification workflow into a practical set of actions enabling the implementation of the FAIR principles, improving the FAIRness evaluation and, consequently, the reuse of (meta)data. Representing a continuous evolution for the FAIRification workflow, this approach can be used as a reference framework. The set of delimited actions is presented below according to each phase of the FAIRification workflow.

**Pre-FAIRification Phase**

The actions established for step 1 (Identify FAIRification Objective) seek to propose a view of the expected results to be achieved through FAIRification. It requires access to data, followed by a preliminary analysis of data and associated metadata. Based on these, it is possible to set goals for the treatment to be performed, identifying the objectives to obtain FAIR (meta)data and defining a set of competency questions that allows it to validate the FAIRification process.

For step 2 (Analyze Data), the actions aim to analyze the data representation according to their format and semantics, the FAIRness evaluation, to check the FAIR maturity level, for example, according to RDA [11], and, finally, to define a relevant subset of the analyzed data for FAIRification.

The actions for step 3 (Analyze Metadata) analyze the metadata associated with the relevant subset of data defined in the previous step and their FAIRness evaluation. It is important to identify the provenance metadata that should be collected for each step of the adapted FAIRification workflow. Figure 2 presents the set of associated actions, for each step of the Pre-FAIRification phase.

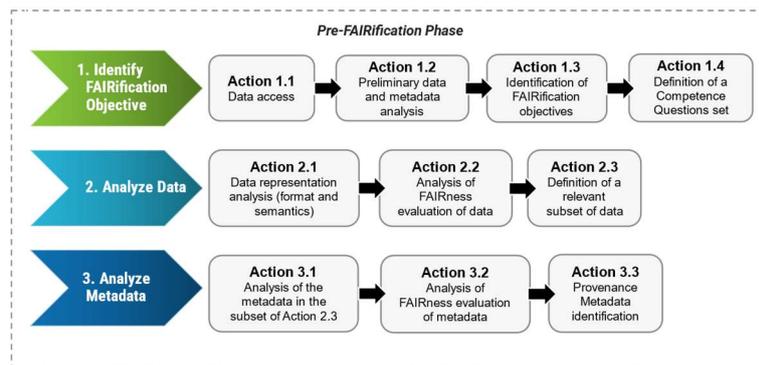

**Fig. 2.** Set of associated actions for each step of the Pre-FAIRification phase.

**FAIRification Phase**

Steps 4a (Define Semantic Data Model) and 4b (Define Semantic Metadata Model) are responsible for the specification of semantic models for data and metadata by identifying and evaluating whether any semantic models already exist and could be reused for them. For cases where no semantic model is available, a new one should be created for the representation of data or metadata.

In steps 5a (Make Data Linkable) and 5b (Make Metadata Linkable), the actions highlight the importance of choosing an RDF framework, as a major step to make (meta)data interoperable and machine-accessible with the association of the semantic models defined in step 4. In step 5b, it is worth mentioning the importance of representing and transforming provenance metadata into a machine-readable and actionable language.

For step 6 (Host FAIR Data and Metadata), the actions make data and metadata available for human and machine use, through various interfaces, such as the adoption of a triplestore for RDF triples and also a FAIR Data Point for metadata storage. The FAIR Data Point adoption facilitates transparent and gradually controlled access over the metadata through four different hierarchical layers: starting with metadata from the FAIR Data Point itself, followed by metadata from the catalog, from the datasets, and from the distributions [12]. Figure 3 presents the set of associated actions for each step of the FAIRification phase.

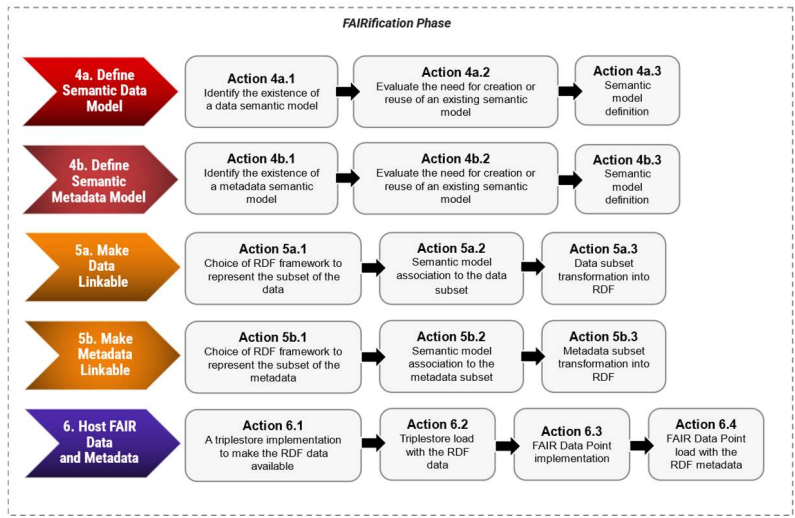

**Fig. 3.** Set of associated actions for each step for the FAIRification phase.

**Post-FAIRification Phase**
Finally, the actions of step 7 (Assess FAIR Data and Metadata) contemplate the assets of the Post-FAIRification process, verifying the objectives and answering the competence questions defined in step 1. Another relevant aspect refers to the assessment of the FAIRness evaluation of data and metadata after the completion of all actions in the adapted FAIRification workflow. Figure 4 shows the set of associated actions for each step in the Post-FAIRification phase.

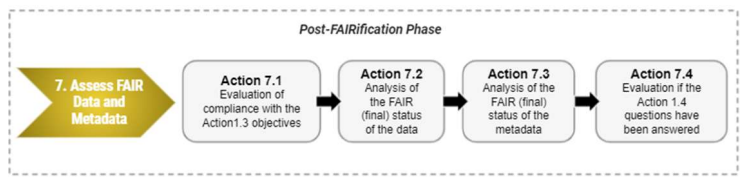

**Fig. 4.** Set of associated actions for each step for the Post-FAIRification phase.

## 4   Solutions to Support the FAIRification Phase

During this study, we investigated solutions capable of supporting and systematizing the FAIRification process, aiming to reduce human errors. The analysis of them helped to understand the recommendations associated with the steps (4a, 4b, 5a, 5b, and 6) of the FAIRification phase. The solutions contributed to validate the actions, promoting the automated support of some steps in the workflow.

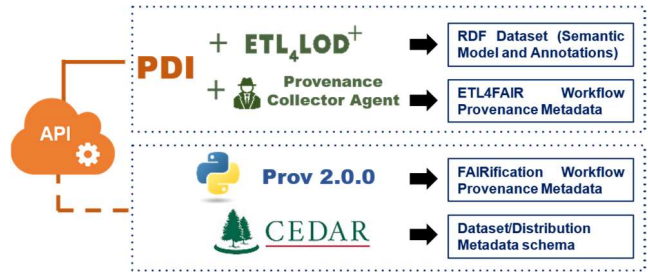

**Fig. 5.** Analyzed solutions and possible integrations.

The solutions experimented and analyzed are presented below, emphasizing their potential to support the steps of the FAIRification phase. Figure 5 presents a summary of the heterogeneous solutions used in the VODAN BR pilot and the possible integrations through their API.

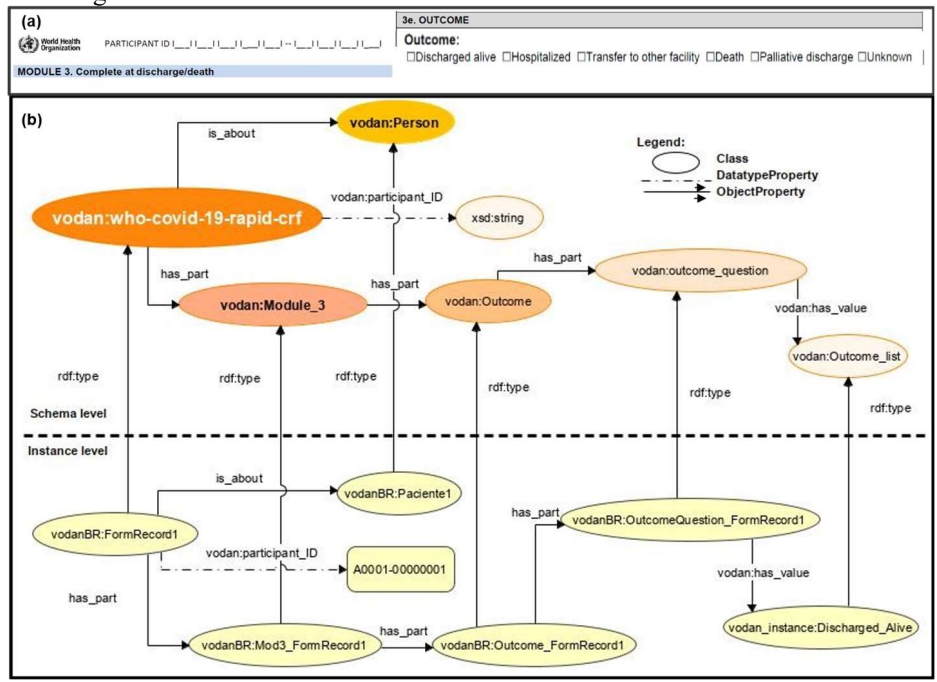

**Fig. 6.** Schematic representation of a patient's outcome using the COVIDCRFRAPID semantics.

For the experiments and analyses of the solutions, we considered the transformation of (meta)data referring to the questions presented in the WHO Case Record Form (CRF-WHO) [13], using the COVIDCRFRAPID semantic model. CRF-WHO was developed by experts to collect relevant anonymous information related to patients with COVID-19. It has three modules: the first one collects the patient's data on the admission day to the health center; the second one, the follow-up, collects daily information such as ICU admission and laboratory results; and the last one summarizes the medical care and collects the outcome information. Figure 6 highlights: (a) the CRF-WHO outcome questions present in "Module3: Complete at discharge/death"; and (b) the semantic model excerpt that handles these questions, associated with instances (example).

### 4.1 ETL4LOD+ Framework

The ETL4LOD+[12] framework provides data cleansing and triplification solutions in the context of Linked Open Data. The framework is an extension of the Pentaho Data Integration (PDI) tool, also known as Kettle, widely used in data ETL processes. This framework provides searching and selecting terms of ontologies and interlinks to other data.

According to our experiments and analyses ETL4LOD+ assists the FAIRification process contributing to the systematization of steps 5a and 5b. Figure 7 shows an example using ETL4LOD+ to transform a patient's outcome data as shown in Figure 6(b). The framework components organize data obtained from the sources into triple (RDF format), according to the respective semantic model, connecting to vocabularies or ontologies (which can be imported).

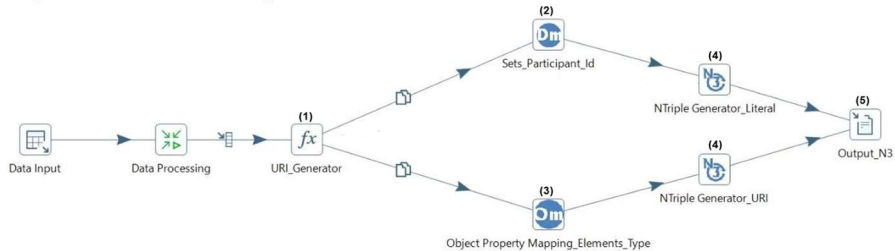

**Fig. 7.** Triplification extract for outcome questions using ETL4LOD+.

To create triplified data, ETL4LOD+ provides several components. In the process depicted in Figure 8, to generate the required URI, we use the "Formula - fx" (1) component. Then, data are annotated with the CRF-OMS ontology. The "Data Property Mapping - Dm" (2) component deals with literal values related to the answers. We also use the "Object Property Mapping - Om" (3) component for other ontology-related items. After mapping, the "NTriple Generator - N3" (4) component serializes the data in N-triples format. Finally, the serialized data is unified in a file through the "File Output" (5) component.

In step 5b, the tool can collect metadata related to the data processing, using solutions such as the Provenance Collector Agent (PCA) (detailed in section 4.2). The associated metadata are obtained in triple format. Furthermore, the tool gathers and organizes other metadata such as the FAIR Data Point dataset and distribution metadata schemas, both generated by the CEDAR tool (detailed in section 4.3), employing the "REST Client" PDI step. In step 6, the tool contributes to the last part of the ETL process, enabling automatic loading of triplified data into a triplestore (currently Virtuoso[13]) or generating an output in files with RDF data serialized in N-Triples. These files can be published in different triplestores.

---

[12] https://github.com/Grupo-GRECO/ETL4LODPlus

### 4.2 ETL4LinkedProv Approach

The purpose of the ETL4LinkedProv[14] approach is to manage the collection and publication of provenance metadata with distinct provenance granularity as Linked Data. The approach uses ETL workflows and employs the Provenance Collector Agent (PCA) component, capturing prospective and retrospective provenance metadata. To support the semantic publication of provenance, ETL4LinkedProv approach uses a set of existing ontologies as PROV-O[15], OPMW[16] and COGS[17] [14].

Through initial analyses and simulation working with the ETL4LOD+ framework, the PCA showed potential to collect provenance metadata associated with an ETL workflow. Our simulation was not as detailed as it was planned due to the version mismatch between ETL4LinkedProv and ETL4LOD+ framework used in the VODAN BR pilot. Therefore, an update of ETL4LinkedProv is currently under development to experiment with the FAIRification process.

As shown in Figure 8, PCA could contribute to steps 5b and 6, capturing prospective and retrospective provenance metadata at different granularity levels and supporting the assessment of the quality and reliability of FAIR provenance metadata. Thus, at the step 6, the captured provenance metadata, as RDF triples semantically annotated using existing provenance ontologies, could be available in a triplestore for SPARQL[18] queries.

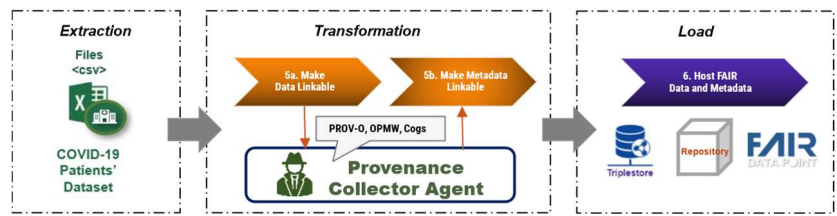

**Fig. 8.** ETL4LinkedProv approach representing the provenance metadata in steps 5a, 5b and 6.

---

[13] https://virtuoso.openlinksw.com/

[14] https://github.com/rogersmendonca/provenance_collector

[15] http://www.w3.org/TR/2013/REC-prov-o-20130430/

[16] https://www.opmw.org/model/OPMW/

[17] http://vocab.deri.ie/cogs

[18] https://www.w3.org/TR/rdf-sparql-query/

### 4.3 Center for Expanded Data Annotation and Retrieval (CEDAR) Workbench

CEDAR Workbench[19] provides a suite of Web-based tools that allows users to build and populate metadata templates, generate high-quality metadata, and share and manage these resources through an API based environment [15]. This solution can assist steps 5b and 6 with respect to metadata schemas established for the FAIR Data Point [12]. Through CEDAR, it is possible to create metadata schemas as templates. These templates must be instantiated with the metadata for the dataset and distribution to be generated. The generated metadata schemas, in RDF N-Quad or JSON-LD, can be accessed directly or through the API and published in the FAIR Data Point. Our experiment used the "REST Client" PDI with CEDAR to collect the metadata schemas of the COVID-19 dataset and its RDF distribution that will be published on the FAIR Data Point.

### 4.4 Prov Python (Prov 2.0.0)

According to our metadata provenance analyses, it was observed the importance of also capturing the provenance of the macro process. For the study of this high-level provenance, the Prov.2.0.0[20] was selected. Prov Python is a library for the implementation of the W3C PROV Data Model[21], with support for serialization in PROV-O (RDF), PROV-XML and PROV-JSON [16].

The experiment and analysis identified a potential solution for steps 5b and 6. In step 5b, the solution provided provenance information, capturing provenance of each step, and it could even collect details of the actions themselves. Figure 9 illustrates a representation of provenance metadata of the step 5b obtained with Prov Python. In step 6, the provenance workflow can be published in a repository or accessed by the ETL4LOD+ tool and joined with the triples of the data file.

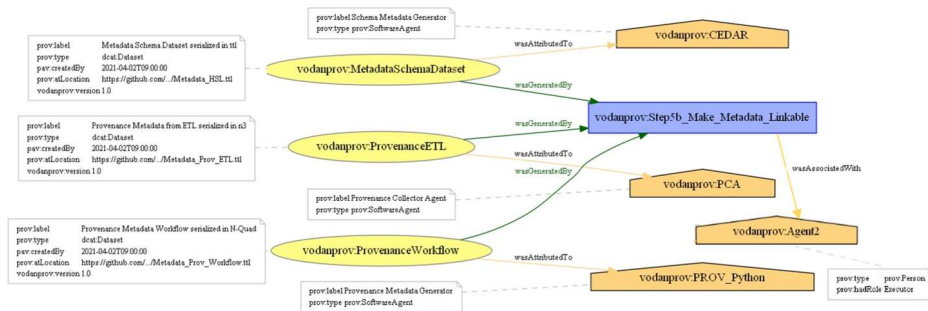

**Fig. 9.** An extract from the provenance metadata for step 5b with Prov Python.

Figure 9, due to space limitations, only shows an extract of the workflow provenance

---

[19] https://metadatacenter.org/
[20] https://openprovenance.org
[21] https://www.w3.org/TR/prov-dm/

metadata based on PROV-O Model, generated by Prov Python. This extract highlights the solutions, represented as software agents (orange pentagons), used in step 5b, the activity (blue rectangle), for metadata generation, and the entities (yellow ovals).

## 5     Discussion and Conclusion

In the VODAN BR pilot, we verified the need for specific actions to be defined in a more detailed FAIRification process, justifying implementation choices of the domain. The rapid spread of the COVID-19 pandemic accelerated the discussion on technology support to publish FAIR (meta)data, although there are still many questions and open issues, especially in the context of metadata management and support. Complementary, it is important to consider a set of best practices from a great number of projects experimenting with the FAIRification process of (meta)data related not only to virus outbreaks, but to life sciences data in general.

This work analyzed the recommendations proposed by the generic workflow for FAIRification and it established an approach inspired by a group of well-delimited actions to support researchers and data stewardship in the generation of FAIR (meta)data. This proposal is being tested in the VODAN BR pilot to guarantee the publication of FAIR data and metadata about COVID-19 hospitalized patients' cases with the support provided by the ETL4FAIR framework. The framework promotes integration between heterogeneous tools to support the process, providing a (semi-)automated workflow for users and reducing error-prone situations.

The first lesson learned along this work is that FAIRification is a complex process in which a multidisciplinary team involvement is extremely important. FAIRification requires several areas of expertise as well as domain knowledge to support each step of the process. Establishing roles and responsibilities for the mapped actions is also important. The second lesson learned is that transforming data and metadata aligned with the FAIR principles is not an easy task. Identifying, choosing, and adapting appropriate data and metadata semantic models are critical actions, as there are many standards disseminated on the Web. Finally, the actions analyzed in this work emphasized the existence of different categories of metadata (for data, data transformation process, and applied process) that can be presented at different granularity levels, contributing to reuse and interoperability. These metadata should be captured throughout a FAIRification process, supported by appropriate tools, whenever possible.

From the exposed context, the importance of establishing actions to define and discuss implementation choices aligned with FAIRification is observed. This contributes to a better organization and maturity of a process that could be assisted by a group of heterogeneous but interoperable solutions. In the near future, we are considering improving the actions proposed in this paper, applying them in different domains.

**Acknowledgements.** This work has been partially supported with students grants from CAPES (Process numbers 223038.014313/2020-19 and 88887.613048/2021-00), CNPq (Process number 158474/2020-1) and UNIRIO.